\documentclass[12pt,draftclsnofoot,journal,onecolumn]{IEEEtran}

\usepackage{cite}
\usepackage{graphicx}
\usepackage{amsmath}
\usepackage{amsfonts}
\usepackage{array}
\usepackage{amssymb}
\usepackage{subfigure}
\usepackage[stable]{footmisc}
\usepackage{multirow}
\newtheorem{lemma}{Lemma}
\newtheorem{theorem}{Theorem}
\newtheorem{corollary}{Corollary}
\newtheorem{remark}{Remark}
\newtheorem{definition}{Definition}
\newtheorem{assumption}{Assumption}

\newtheorem{property}{Property}
\newtheorem{example}{Example}

\begin{document}
\title{Fundamentals of Inter-cell Overhead Signaling in Heterogeneous Cellular Networks
}
\author{Ping Xia, Han-Shin Jo and Jeffrey G. Andrews\\
\thanks{The authors are with the Wireless Networking and Communications Group in the Department of Electrical and Computer Engineering at The
University of Texas at Austin. Email: pxia@mail.utexas.edu, han-shin.jo@austin.utexas.edu and jandrews@ece.utexas.edu. Dr. Andrews is the contact author. This research was supported by Motorola Solutions and the National Science Foundation. Manuscript last modified: \today}
}
\maketitle
\begin{abstract}
Heterogeneous base stations (e.g. picocells, microcells, femtocells and distributed antennas) will become increasingly essential for cellular network capacity and coverage. Up until now, little basic research has been done on the fundamentals of managing so much infrastructure -- much of it unplanned -- together with the carefully planned macro-cellular network. Inter-cell coordination is in principle an effective way of ensuring different infrastructure components behave in a way that increases, rather than decreases, the key quality of service (QoS) metrics. The success of such coordination depends heavily on how the overhead is shared, and the rate and delay of the overhead sharing. We develop a novel framework to quantify overhead signaling for inter-cell coordination, which is usually ignored in traditional $1$-tier networks, and assumes even more importance in multi-tier heterogeneous cellular networks (HCNs). We derive the \emph{overhead quality contour} for general $K$-tier HCNs -- the achievable set of overhead packet rate, size, delay and outage probability -- in closed-form expressions or computable integrals under general assumptions on overhead arrivals and different overhead signaling methods (\emph{backhaul} and/or \emph{wireless}). The \emph{overhead quality contour} is further simplified for two widely used models of overhead arrivals: \emph{Poisson} and \emph{deterministic} arrival process. This framework can be used in the design and evaluation of any inter-cell coordination scheme. It also provides design insights on backhaul and wireless overhead channels to handle specific overhead signaling requirements.
\end{abstract}

\section{Introduction}
Heterogeneous cellular networks (HCNs) -- comprising macro base stations (BSs) and overlaid infrastructure (e.g. picocells, femtocells and distributed antennas) -- have recently emerged as a flexible and cost-effective way of handling the exploding and uneven wireless data traffic demands, which are expected to increase indefinitely\cite{VC08CommMag,Qcomm10,TROPOS07}. By improving network capacity and coverage with significantly lower capital and operational expenses, such networks are gaining industrial momentum as both a short-term tactic and a long-term strategy.

\subsection{Inter-cell Coordination Techniques in HCNs}
The management of HCNs is significantly more difficult than the traditional $1$-tier macrocell case, which is already considered challenging. The different kinds of BSs have distinct spatial densities, transmit powers, cell sizes, and backhaul capabilities. Further, the overlaid infrastructure will often be added over time in ad hoc locations\cite{VC08CommMag,Qcomm10,TROPOS07,ZhaAnd08}. Centralized control of all these BSs involves a potentially enormous amount of overhead messaging and is considered impractical. Decentralized inter-cell coordination is in principle an effective way of organizing HCNs for coordinated multipoint (COMP), cooperative scheduling and handoffs. In general, inter-cell coordination enables neighboring cells to successfully co-exist and allows cooperative gains\cite{Gesbertetal10}, which includes improvements to signal-to-interference-plus-noise ratio (SINR), spectral efficiency and/or outage rates.

Many coordination techniques are shown to have large cooperative gains in theory. However, the assessment of these gains usually ignores the inherent cost of overhead sharing: the overhead (e.g. CSI and user scheduling) is shared at limited rate with quantization error and delay\cite{SR09CSI,SR09Asilomar}. Practical concerns on overhead lead to non-trivial gaps between real and theoretical cooperative gains. An example is downlink joint processing COMP in the $1$-tier case, which ideally introduces a multi-fold throughput improvement\cite{ShaZai01,SomZaiSha07,Gesbertetal10}. However, industrial simulations and field trails show that real throughput gain is disappointing -- less than $20\%$ -- and the major limiting factor is sharing CSI and other overhead among cells\cite{SR09CSI,SR09Asilomar,Qcomm10CTW,Irmeretal11CommMag}. Mathematically, the achievable gain is a function of overhead parameters: 1) $\mathcal{T}$, the overhead packet interarrival time (the inverse of which is overhead packet rate); 2) $B$, the overhead packet bit size; and 3) $\mathcal{D}$, the overhead delay. It is therefore important to evaluate cooperative gains in terms of the achievable values of these overhead signaling parameters.

\subsection{Previous Models for the Overhead Parameters}
The model of limited overhead bit rate, which is the product of overhead packet rate $1/\mathbb{E}[\mathcal{T}]$ and packet size $B$, is previously considered for wireless overhead signaling\cite{Loveetal08}. It is not considered for backhaul signaling in the traditional $1$-tier macrocell case (except that overhead includes user data\cite{Somekhetal08,SanSomPooSha09}), assuming macro BSs are equipped with high capacity backhaul. However, it is not always the case for BSs in HCNs. In particular, femtocells must leverage third-party IP based backhaul (e.g. DSL and cable modem) that is aggregated by a gateway and so has much lower rate\cite{VC08CommMag,FemtoBackhaul}.

Besides average rate, the natural dynamics in overhead interarrival time $\mathcal{T}$ are often ignored. In coordination techniques where inter-cell overhead is driven or influenced by unplanned incidents (e.g. during inter-cell handoffs overhead is generated when a user crosses cell boundaries), the interarrival time $\mathcal{T}$ varies over time. However, previous works simply assume $\mathcal{T}$ as a constant value (e.g. several symbol time\cite{ZhaKouAndHea11Multimode}).

Perhaps the most important piece missing from previous works is an appropriate model on overhead delay $\mathcal{D}$ in general multi-tier HCNs. In $1$-tier macrocell case, the backhaul interface between neighboring BSs is modeled as nearly delay-free\cite{ShaZai01,FosKarVal06,SomZaiSha07,Gesbertetal10}. This assumption may hold if macrocells are directly interconnected by high speed Ethernet\cite{Irmeretal11CommMag}, but is far from reality in most network configurations\cite{SR09CSI,SR09Asilomar,HuaWei09}. More than likely, it is not applicable to overlaid BSs with generally lower capacities and more complicated protocols\cite{VC08CommMag,FemtoBackhaul}. For wireless signaling (e.g. to-be-defined overhead channels in LTE-A), the overhead delay is also very different from the $1$-tier case due to distinct statistics of spatial interference in HCNs\cite{FleStoSim97,AndBacGan10,DhiGanBacAnd11,JoSanXiaAnd11}. With even moderate mobility, delay in side information results in an irreducible performance bound that cannot be overcome even with much higher rate and more frequent overhead messages\cite{AliTse10}.

In short, the appropriate models on overhead parameters in multi-tier HCNs are currently missing but of critical importance for the design and evaluation of coordination techniques. It is thus desirable to develop a general framework to quantify the feasible set of overhead parameters $(\mathcal{T}, B, \mathcal{D})$ as a function of various HCNs setups, rather than heuristically for each possible network realization.

\subsection{Contributions}
In Section II, we develop general models for the overhead parameters in HCNs: 1) a Gamma distribution model on overhead interarrival time $\mathcal{T}$, which contains two important and opposite special cases: \emph{deterministic} and \emph{Poisson} overhead arrivals; 2) queuing models on backhaul servers (e.g. switches, routers and gateways) to characterize backhaul overhead delay $\mathcal{D}$; and 3) a stochastic geometry model on HCN spatial interference to characterize wireless overhead delay $\mathcal{D}$.

From such models, we propose a novel framework \emph{overhead quality contour} to quantify feasible overhead parameters $(\mathcal{T},B,\mathcal{D})$ as a function of overhead channel realizations and overhead arrivals. We derive its general expressions in computable integrals for backhaul (Section III) and wireless overhead signaling (Section IV), which are simplified to closed-form results in two widely assumed overhead arrivals: \emph{deterministic} and \emph{Poisson}. We show mathematically and through numerical simulations that previous models, compared with our framework, are over-optimistic about achievable overhead rate, delay and outage probability, which explains the non-trivial gaps between their predictions and the real cooperative gains.

The \emph{overhead quality contour} can be leveraged for the following general purposes.

\textbf{The Evaluation and Optimization of HCN Coordinations.} The \emph{overhead quality contour} can be directly used for the analysis of specific HCN coordination techniques by determining: 1) the feasibility of these techniques, i.e. if their overhead requirements (e.g. overhead outage below some threshold) lie in the \emph{overhead quality contour}; 2) if feasible, their possible overhead signaling options, i.e. achievable set of $(\mathcal{T},B,\mathcal{D})$ in different overhead signaling methods (backhaul and/or wireless). The gains of proposed coordination techniques can then be maximized by choosing the appropriate overhead signaling option.

\textbf{The Design of HCN Overhead Channels.} During the deployment of HCNs, the proposed framework is also useful in providing design insights on overhead channel setups to facilitate inter-cell coordinations. Based on the {overhead quality contour}, we derive tight lower bound in Section III on backhaul servers' rate as a function of overhead signaling requirements and backhaul connection scenarios (i.e. the number of backhaul servers). Similarly, the lower bound on wireless overhead channel bandwidth is characterized in Section IV. The optimal setups to achieve these lower bounds are also identified.

\section{System Model}
A heterogeneous cellular network -- comprising $K$ types of base stations (BSs) with distinct spatial densities and transmit powers -- can be modeled as a $K$-tier network, with $\mathcal{B}_k$ denoting the set of BSs in the $k^{th}$ tier. For example, high-power macrocells overlaid with denser and lower power femtocells are referred as two-tier femto networks\cite{ChaAnd09UL}. The locations of BSs (e.g. pico and femto BSs) in each tier can be modeled by an appropriate spatial random process, since their locations are usually unplanned. Surprisingly, it is also a reasonable model for HCNs including macro BSs, providing as much accuracy as the widely used grid model as compared to a real BS deployment\cite{FleStoSim97,AndBacGan10,DhiGanBacAnd11,JoSanXiaAnd11}. Therefore, we assume all tiers are independently distributed on the plane $\mathbb{R}^2$ and BSs in $\mathcal{B}_k$ are distributed according to Poisson Point Process (PPP) $\Phi_k$ with intensity $\lambda_k$. Note that this assumption only affects the SINR characterization of the wireless overhead channel (i.e. its CDF $q\{\cdot\}$ in Lemma \ref{LemmaSIR}), while our results on overhead signaling hold under various SINR distributions.

In a $K$-tier network, a base station $\mathrm{BS}_0$ intends to coordinate with its neighboring base station $\mathrm{BS}_n$, from whom it receives the strongest long-term average power (which means strongest interference if not coordinated). Therefore, $\mathrm{BS}_0$ needs to constantly know the key parameters of $\mathrm{BS}_n$, such as its user scheduling and/or the scheduled user's CSI. Suppose during each overhead signaling slot, $\mathrm{BS}_n$ compresses these parameters (e.g. by quantization and coding) into an overhead packet of $B$ bits and transmits it to $\mathrm{BS}_0$. To quantify the feasible set of overhead parameters $(\mathcal{T}, B, \mathcal{D})$, we describe the models on overhead interarrival time $\mathcal{T}$ and delay $\mathcal{D}$ in the following.

\subsection{Overhead Message Interarrival Time}
\begin{assumption} \label{ASAverageStateTime}
\emph{The overhead arrival is assumed to be a stationary homogeneous arrival process with packet rate $\eta$, i.e. the packet interarrival times have the same distribution with $\mathbb{E} \left[\mathcal{T}\right]= 1/\eta$.}
\end{assumption}
At its most general, we assume the interarrival time is gamma distributed with parameter $M$
\begin{align} \label{EquTmodel}
\mathcal{T}^{(M)} \sim \mbox{Gamma} \left(M, \frac{1}{M\eta}\right).
\end{align}
For various values of $M$, the average interarrival time is still $\mathbb{E} \left[\mathcal{T}^{(M)}\right] = 1/\eta$. This model of $\mathcal{T}$ includes two widely used models on overhead arrivals as special cases: deterministic and Poisson arrivals.

\textbf{Deterministic Overhead Arrivals.} The interarrival time $\mathcal{T}$ can be a constant determined by $\mathrm{BS}_0$ and $\mathrm{BS}_n$ based on standards or other agreements. An example is joint frequency allocation in LTE: base stations utilize certain preamble bits in each frame as their coordination message, to specify the frequency allocations for their users' data in this frame. Therefore the overhead message is generated in every $10$ ms (i.e. each LTE frame)\cite{GhoZhaAndMuh10LTE}. In (\ref{EquTmodel}), $M \rightarrow \infty$ gives constant interarrival time \begin{align} \label{EquACinterval}
\mathcal{T}^{(M)} \stackrel{d.}{\rightarrow} \mathcal{T} =1/\eta,
\end{align}
where $\stackrel{d.}{\rightarrow}$ means convergence in distribution.

\textbf{Poisson Overhead Arrivals.} The interarrival time $\mathcal{T}$ can also be random, determined by the users or other cells rather than $\mathrm{BS}_0$ and $\mathrm{BS}_n$ themselves. An example is user cell associations. As the users roam around, they choose their serving cells based on certain metrics including received power and congestion. Such choices will change the cell parameters (e.g. user scheduling and resource allocations) at $\mathrm{BS}_n$, which means a new overhead message should be generated and shared with $\mathrm{BS}_0$. The overhead arrivals are thus random and often modeled as Poisson process with exponential interarrival time
\begin{align}
\mathcal{T} \sim \exp (\eta).
\end{align}
It is known that exponential distribution is also a special case of (\ref{EquTmodel}) with $M=1$.

These two special cases of practical interest provide insights into two opposite extremes since for a given rate, deterministic arrivals are the least random while Poisson arrivals are the most random (maximum entropy). An arbitrary overhead arrival model is therefore bounded by these two extreme cases (which also have practical significance).

\subsection{Overhead Delay in Backhaul Network}
\begin{figure}[hp]
\centerline{
\includegraphics[width=3.5in]{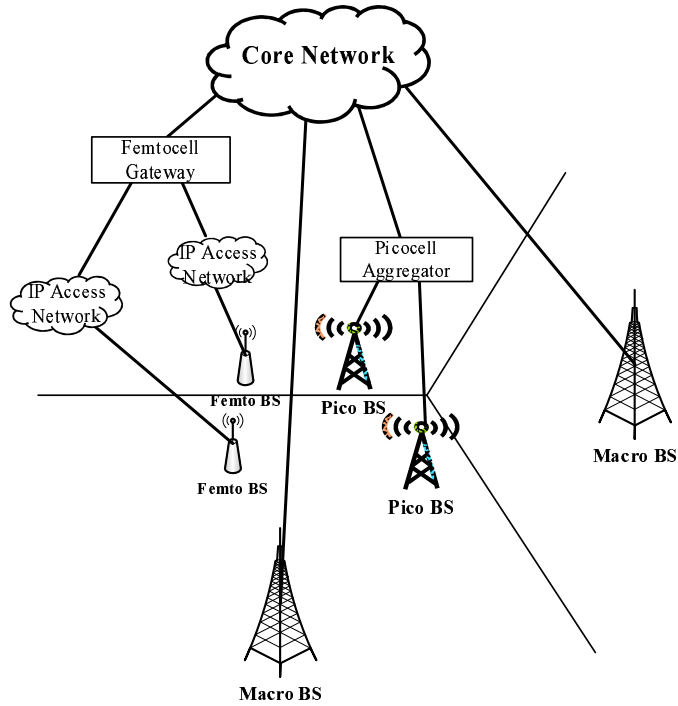}}
\caption{The base station locations and backhaul deployments of a $3$-tier heterogeneous cellular network, comprising for example macro (tier 1), pico (tier 2) and femto (tier 3) BSs.} \label{PicSystemModel}
\end{figure}

In HCNs, the backhaul connection between BSs are likely to be diverse as shown in
Fig. \ref{PicSystemModel}. In general, backhaul overhead delay comprises two parts: 1) latencies from switches, routers and gateways (generally termed \emph{backhaul servers}); and 2) the transmission delay of the wire (e.g. fiber lines and copper wires) or wireless links (e.g. microwave). The latter kind of latency is quite small and often neglected. For example, as the backhaul path between clustered picocells or co-located BSs contains few servers, the backhaul delay can be as low as $1$ ms\cite{LTE9,HuaWei09,VenKul08WirelessBackhaul}. The backhaul network can therefore be reasonably modeled as $N$ tandem servers.
\begin{assumption} \label{ASPacketDrop}
\emph{A backhaul server drops overhead message(s) in its system upon the arrival of new overhead, i.e. overhead messages do not queue at any backhaul servers.}
\end{assumption}
In this paper, we do not assume any retransmission for overhead packets due to their time sensitivity. Therefore once observing new overhead arrivals, the backhaul servers know that existing overhead packet are outdated and should be dropped.

\begin{assumption} \label{ASServerDelay}
\emph{We assume the backhaul servers have exponential service time, the $i^{th}$ of which allocates bit rate $\mu_i$ to overhead packets.}
\end{assumption}
Note that the parameters $\{\mu_i\}$ in Assumption \ref{ASServerDelay} are dependent on the scheduling policies of backhaul servers. In the following, we list a few common examples.
\begin{example}
\emph{(Pre-emptive Scheduling)}: In this case, servers recognize the extreme delay sensitivity of overhead packets and identify them as the highest priority traffic. Thus, overhead will be served before all other traffic in a pre-emptive way\cite{WerWanSyn07} and its allocated rate $\mu_i$ is indeed the total service rate $\mu_i^{\textrm{total}}$.
\end{example}
\begin{example}
\emph{(High Priority Scheduling)}: Servers identify overhead as a real-time flow with stringent delay and serve them before packets with an elastic delay requirement (e.g. non-real-time traffic such as web surfing)\cite{SadmadSam09}. Suppose other real time traffic is Poisson with total rate $\nu^{\textrm{rt}}$, the bit rate experienced by overhead packets will be $\mu_i = \mu_i^{\textrm{total}}-\nu^{\textrm{rt}}$.
\end{example}
\begin{example}
\emph{(Equal Priority Scheduling)}: All traffic is scheduled with equal priority at the servers. This is close to the worst case since the delay sensitivity of overhead traffic is ignored\cite{SadmadSam09}. Suppose the data traffic are Poisson with rate $\nu^{\textrm{d}}$, we then have $\mu_i= \mu_i^{\textrm{total}}-\nu^{\textrm{d}}$.
\end{example}

Under Assumption \ref{ASPacketDrop} and \ref{ASServerDelay}, the overhead delay from the $i^{th}$ backhaul server is
\begin{align}
\mathcal{D}_i \sim \exp \left(\frac{\mu_i}{B}\right),
\end{align}
where $\mu_i$ is the effective bit rate and $\frac{\mu_i}{B}$ is thus overhead packets rate per second. For overhead messages not dropped during transmission, the end-to-end backhaul delay is
\begin{align} \label{EquBackhaulLatency}
\mathcal{D} = \sum\limits_{i=1}^N \mathcal{D}_i.
\end{align}
The values of $N$ and $\frac{\mu_i}{B}$ in (\ref{EquBackhaulLatency}) depend on the specific backhaul configurations between $\mathrm{BS}_n$ and $\mathrm{BS}_0$. For the backhaul connection between macro BSs, $N$ is typically around $10$ to $20$ and $\mu_i/B$ is thousands of packets per second\cite{HuaWei09}.
In the most general case, the cumulative distribution function (CDF) of delay $\mathcal{D}$ is very complicated and still under investigation \cite{SA97,SF10}. In our paper, we consider a scenario of practical interest: $\mu_i \neq \mu_j $ for any $i \neq j$. The CDF is then
\begin{align} \label{EqubackhaulCDF}
\mathcal{F}(d,B, \{\mu_i\}|_{i=1}^N) \equiv \mathbb{P} (\mathcal{D} \leq d) =
\sum\limits_{i=1}^{N} a_i (1- e^{-\mu_i d/B}), \quad \textrm{where } a_i=\prod\limits_{j\neq i} \frac{\mu_j}{\mu_j-\mu_i}\,.
\end{align}
We now derive an important property of $\{a_i\}$ in below, which will be frequently used in the sequel.
\begin{property} \label{PropertySum}
$\sum_{i=1}^N a_i \frac{\mu_i}{\mu_i+x} = \prod_{i=1}^N \frac{\mu_i}{\mu_i+x}$, $\forall x \geq 0$.
\end{property}
\begin{IEEEproof}
See Appendix \ref{ProofPropertySum}.
\end{IEEEproof}

In this subsection the overhead delay in backhaul signaling is modeled and its CDF $\mathcal{F} (\cdot)$ is also derived. In the following, we characterize wireless overhead delay by using our SIR results\cite{JoSanXiaAnd11} in $K$-tier HCNs.

\subsection{Overhead Delay in Wireless Overhead Channel\footnote{It is important to clarify the fundamental differences of wireless backhaul (e.g. connect BSs to core network through microwave) and wireless overhead channel. In the former case, the microwave link is interference free and has large capacity\cite{VenKul08WirelessBackhaul}, but overhead will be routed in the backhaul network. In the latter case, overhead is directly shared between $\mathrm{BS}_0$ and $\mathrm{BS}_n$ without routing, but the wireless overhead channel has much lower capacity (because of interference and often constrained bandwidth).}}
The wireless channel can be modeled as
\begin{align}
h (x) = g L |x|^{-\alpha},
\end{align}
where $g$ is the short-term fading, $L$ is the wall penetration loss (e.g. femtocells are usually deployed indoors), $x$ is the Euclidian distance between transmitter and receiver, and $\alpha$ is the path loss exponent. In this paper, we consider i.i.d. Rayleigh fading with unit mean power, i.e. $g \sim \exp(1)$. Denote $P_k$ as the transmitting power of BSs in the $k^{th}$ tier while $\alpha_k$ and $L_k$ as path loss exponent and wall penetration of their channels to $\mathrm{BS}_0$.

As mentioned before, $\mathrm{BS}_0$ chooses to coordinate with $\mathrm{BS}_n$ if it receives the strongest long-term (i.e. with fading averaged out) power from $\mathrm{BS}_n$. The received signal-to-interference ratio (SIR)\footnote{Conventional cellular networks are generally interference-limited while thermal noise is negligible. Interference is even more significant in HCNs due to the overlaid BSs, generally of high density. Therefore, in this paper we neglect thermal noise.} at $\mathrm{BS}_0$ under this model is derived in following lemma. See Lemma $1$ and $2$ and Theorem $1$ in \cite{JoSanXiaAnd11} for proof.
\begin{lemma} \label{LemmaSIR}
\cite{JoSanXiaAnd11} \emph{The probability that $\mathrm{BS}_n$ associates with $k^{th}$ tier is}
\begin{align}
\mathbb{P} (\mathrm{BS}_n \in \mathcal{B}_{k}) = 2\pi \lambda_k \int_{x>0} x \exp\left(-\pi\sum_{j =1}^K \lambda_j  \,\hat{P}_{j}^{2/\alpha_j} \, x^{2/\hat{\alpha}_{j}} \right) \mathrm{d} x.
\end{align}
\emph{Thus, the received SIR at $\mathrm{BS}_0$ is}
\begin{align} \label{EquSIRCDF}
q_k\{\beta\} & \stackrel{\Delta}{=} \mathbb{P} (\mbox{SIR} \leq \beta | \mathrm{BS}_n \in \mathcal{B}_{k} ) \nonumber \\
&= 1-\frac{2\pi\lambda_k}{\mathbb{P} (\mathrm{BS}_n \in \mathcal{B}_{k})} \int_{x>0} x \,\exp\left(-\pi\sum\limits_{j=1}^K \lambda_j  \hat{P}_{j}^{2/\alpha_j} (1+\mathcal{Z} (\beta,\alpha_j))x^{2 /\hat{\alpha}_{j}} \,\right)  \mathrm{d}x,
\end{align}
\emph{where $\hat{P}_j=\frac{P_j L_j}{P_k L_k}$ and $\hat{\alpha}_j=\frac{\alpha_j}{\alpha_k}$ and }
\begin{align}
\mathcal{Z} (\beta, \alpha_j) = \beta^{\frac{2}{\alpha_j}} \int_{\beta^{-\frac{2}{\alpha_j}}}^\infty \frac{1}{1+ u^{\alpha_j/2}} \mathrm{d} u.
\end{align}
\end{lemma}
The function $q_k\{\cdot\}$ in Lemma \ref{LemmaSIR} is expressed in its most general form and can be significantly simplified. For example, $q_k\{\beta\}=1-\frac{1}{1+\mathcal{Z}(\beta,\alpha)}$  when all path loss exponents are the same (See Corollary $2$ in \cite{JoSanXiaAnd11}).

Similar to Assumption \ref{ASPacketDrop} for backhaul signaling, $\mathrm{BS}_n$ can be reasonably assumed to drop existing overhead packets upon the arrival of new overhead. The overhead packets, if not dropped during transmission, therefore experience delay given by
\begin{align} \label{EquWirelessLatency}
\mathcal{D}= \frac{B}{W \log (1 + \mbox{SIR})},
\end{align}
where $B$ is the overhead packet size, $W$ is the overhead channel bandwidth and the distribution of SIR is given in (\ref{EquSIRCDF}).

\subsection{Fundamental Evaluation Metric}
With overhead interarrival time $\mathcal{T}$ and delay $\mathcal{D}$ modeled, we here define overhead outage $p_e$.
\begin{definition} \label{DefOutage}
\emph{An overhead message is successful if it arrives at the destination $\mathrm{BS}_0$ before being outdated (i.e $\mathcal{D} \leq \mathcal{T}$, since an overhead is not outdated until a new one is generated) and before a hard deadline $d$ (i.e. $\mathcal{D} \leq d$). Otherwise, it is defined as in outage.}
\end{definition}
The outage defined above is the probability that an overhead block is not fully received before a certain deadline specified by the coordination techniques. It is indeed the overhead block error, not including the effect of coding and complicated overhead transmission schemes\cite{AliTse10}. Based on Definition \ref{DefOutage}, the fundamental evaluation metric of this paper -- the \emph{overhead quality contour} is thus defined as
\begin{align} \label{DefContour}
\mathcal{Q}_o \stackrel{\triangle} {=} \left\{\left(\mathcal{T},B, d, p_e \right): \; p_e  = 1-\mathbb{P} (\mathcal{D} \leq \mathcal{T}, \mathcal{D} \leq d) \right\},
\end{align}
where $\mathcal{T}$ is the overhead interarrival time, $B$ is the overhead packet size, $d$ is the required overhead deadline (i.e. maximal tolerable delay), and $p_e $ is the corresponding outage probability. Note that the delay $\mathcal{D}$ is fully characterized by $d$ and $p_e$, and thus is not explicitly included in $\mathcal{Q}_o$.

This metric above determines the feasible set of overhead parameters $\{\mathcal{T},B,d,p_e\}$ as a function of overhead signaling configurations in HCNs (e.g. overhead arrival process and channel parameters). It can be used to identify the feasibility of various coordination techniques (i.e. if their overhead requirements lie in the \emph{overhead quality contour}) and quantify their possible overhead signaling options. It also provides insights on overhead channel configurations to handle overhead signaling as required by specific HCN coordination techniques. In short, as will be illustrated in Section III and IV, this framework can be leveraged for the evaluation and design of coordination techniques and HCN overhead channel setups.

\section{Overhead Quality Contour in Backhaul Signaling}
This section presents the main results for backhaul overhead signaling. The \emph{overhead quality contour} is quantified when $\mathrm{BS}_n$ and $
\mathrm{BS}_0$ share overhead through their dedicated backhaul. The backhaul network is in general modeled as $N$ tandem servers with overhead packet processing rate $\{\mu_1/B,\ldots, \mu_N/B\}$. The specific backhaul configurations (i.e. the values of $N$ and $\{\mu_i\}$) are heavily contingent on the types of $\mathrm{BS}_0$ and $\mathrm{BS}_n$, which we will discuss in detail in the numerical results. We first consider the general case of overhead arrivals with gamma distributed interarrival time. We then focus on the two special cases previously identified: 1) deterministic overhead arrivals; and 2) Poisson overhead arrivals.
\subsection{General Case and Main Results}
\begin{theorem} \label{ThmBackhual}
\emph{For backhaul overhead signaling between $\mathrm{BS}_n$ and $\mathrm{BS}_0$ with interarrival time $\mathcal{T} \sim $ \emph{Gamma} $\left(M,\frac{1}{M\eta}\right)$, the overhead quality contour is}
\begin{align} \label{EquThmBackhaul}
\mathcal{Q}_o = \left\{(\mathcal{T},B, d,p_e): p_e =\sum\limits_{i=1}^N a_i \left[\left(1-\frac{\gamma(M,M\eta d)}{\Gamma(M)}\right)e^{-\frac{\mu_i d}{B}}+\frac{\gamma\left(M,M\eta d+\frac{\mu_i d}{B}\right)}{\Gamma(M)} \left(\frac{M\eta}{M\eta+\frac{\mu_i}{B}}\right)^M\right] \right\},
\end{align}
\emph{where $\{a_i\}$ are defined in (\ref{EqubackhaulCDF}), $\Gamma(\cdot)$ is the gamma function and $\gamma(M,x)$ is the lower incomplete gamma function given by}
\begin{align}
\gamma(M,x) = \int_0^x t^{M-1} e^{-t} \mathrm{d} t.
\end{align}
\end{theorem}
\begin{IEEEproof}
See Appendix \ref{ProofThmBackhaul}.
\end{IEEEproof}

Theorem \ref{ThmBackhual} quantifies all plausible overhead parameters that can be supported by given backhaul configurations. Since many coordination techniques often have additional requirements on several overhead parameters (e.g. requiring $p_e \leq 0.1$), their feasible overhead sets are strict subsets of $\mathcal{Q}_o$. In theory, these subsets can be determined from Theorem \ref{ThmBackhual} by, for example, restricting $p_e \leq 0.1$ in (\ref{EquThmBackhaul}). However, it is computationally hard in practice to derive feasible set of $(\mathcal{T},B,d)$ under a given outage requirement. In the following, we derive simpler bounds on (\ref{EquThmBackhaul}), which can be easily used to characterize the feasible set of several overhead parameters given others.

According to its definition and observations from Theorem \ref{ThmBackhual}, the outage probability $p_e$ is an increasing function on the overhead rate $\eta$ while a decreasing function on the deadline requirement $d$. For example, the outage probability is zero when $\eta \rightarrow 0$ and $d \rightarrow \infty$ as shown in Theorem \ref{ThmBackhual}. Therefore, it has the following two lower bounds.
\\
\textbf{Lower Overhead Rate:} By letting the overhead packet rate $\eta$ go to zero, overhead packets have very long lifetimes (i.e. $\mathbb{E}[\mathcal{T}]=1/\eta \rightarrow \infty$) and overhead outage only comes from the probability of not meeting the hard deadline (i.e. $\mathcal{D} > d$).
\begin{align} \label{EquBackhaulepsilonlb1}
p_e \geq  \sum_{i=1}^N a_i e^{-\mu_i d/B} =1-\mathcal{F} (d, B, \{\mu_i\}|_{i=1}^N) \stackrel{\triangle}{=}p_e^{lb,1}.
\end{align}
\textbf{Relaxed Delay Deadline:} By letting delay deadline $d$ go to infinity, the overhead delay deadline is relaxed and outage only comes from the probability of being outdated during transmission (i.e. $\mathcal{D} > \mathcal{T}$).
\begin{align} \label{EquBackhaulepsilonlb2}
p_e \geq \sum\limits_{i=1}^N a_i \left(\frac{M\eta}{M\eta+\mu_i/B}\right)^M \stackrel{\triangle}{=}p_e^{lb,2}.
\end{align}

\begin{remark} \label{ReLBGeneral}
\emph{For backhaul signaling, a lower bound for overhead outage probability is}
\begin{align} \label{EquLBGeneral}
p_e \geq \max \left(p_e^{lb,1}, p_e^{lb,2}\right),
\end{align}
\emph{where $p_e^{lb,1}$ and $p_e^{lb,2}$ are given in (\ref{EquBackhaulepsilonlb1}) and (\ref{EquBackhaulepsilonlb2}) respectively.}
\end{remark}
The lower bound in Remark \ref{ReLBGeneral} is achieved under deterministic arrivals (stated in Remark \ref{ReLBAC}) and fairly tight under general arrivals (shown in Section V). With much simplified but still accurate results, Remark \ref{ReLBGeneral} can be leveraged to estimate feasible overhead sets for various coordination techniques. For example, for coordination techniques requiring $p_e \leq 0.1$, the feasible set of $(\mathcal{T},B,d)$ can be easily determined by solving $\max \left(p_e^{lb,1}, p_e^{lb,2}\right) \leq 0.1$.

It is interesting to compare feasible overhead parameters $(\mathcal{T},B,\mathcal{D})$ quantified by our framework with previous works. Without the model capturing the randomness in interarrival time, $\mathcal{T}$ is implicitly assumed as the constant $1/\eta$. Overhead backhaul delay $\mathcal{D}$ in (\ref{EquBackhaulLatency}) is often neglected or simply assumed as the average latency
\begin{align}
\mathcal{D} =\mathbb{E} \left[\sum_{i=1}^N \mathcal{D}_i\right] = \sum_{i=1}^N \frac{B}{\mu_i}\, .
\end{align}
Under the above simplified models, the \emph{overhead quality contour} will reduce to
\begin{align} \label{EquFakeBackhaul}
\mathcal{Q}_o = \left\{(\mathcal{T},B,d, p_e): p_e = 1- \mathbb{P} (\mathcal{D} \leq \mathcal{T}, \mathcal{D} \leq d) = 1- \mathbf{1}\left(\sum_{i=1}^N \frac{B}{\mu_i} \leq \min(d,1/\eta)\right)\right\}.
\end{align}
Obviously the feasible overhead parameters defined in (\ref{EquFakeBackhaul}) are vastly different from (\ref{EquThmBackhaul}). For example, under given backhaul servers' rates $\{\mu_i\}$, overhead outage in (\ref{EquFakeBackhaul}) can be zero under finite values of $d$ and $1/\eta$. However, the lower bound in Remark \ref{ReLBGeneral} shows that $p_e=0$ \emph{iff} $d \rightarrow \infty$ \emph{and} $1/\eta \rightarrow \infty$. In short, the natural randomness in $\mathcal{T}$ and $\mathcal{D}$ crucially determines the feasible overhead signaling contours and will be discussed more in Section V.

In general, the \emph{overhead quality contour} $\mathcal{Q}_o$ can be leveraged for the design and evaluation of coordination techniques in HCNs. For example, in below we provide backhaul design guidelines to effectively support overhead signaling required by coordination techniques.

\begin{corollary} \label{CoroBackhaulInsights}
\emph{For a given overhead requirement $(\mathcal{T},B,d,p_e)$ from coordination techniques, the backhaul configuration, i.e. the values of $N$ and $\{\mu_i\}$, must satisfy the following inequalities}
\begin{align} \label{EquMuBound}
\bar{\mu} & \geq \frac{B}{d}  \,\gamma^{-1} \left((1-p_e) (N-1)!, N\right)\, , \\
\bar{\mu} & \geq \frac{M \eta B\sqrt[N]{1-p_e}} {\sqrt[N]{\binom{M+N-1}{N}}-\sqrt[N]{1-p_e}} \, ,
\end{align}
\emph{where $\bar{\mu}$ is the average service rate $\bar{\mu} = \frac{\sum_{i=1}^N \mu_i}{N}$ and $\gamma^{-1} (x,N)$ is the inverse incomplete gamma function given by}
\begin{align}
x= \gamma(N,y) = \int_{0}^y t^{N-1}e^{-t} \mathrm{d}t \; \Leftrightarrow\; y=\gamma^{-1} (x,N).
\end{align}
\end{corollary}
\begin{IEEEproof}
See Appendix \ref{ProofCoroBackhaulInsights}.
\end{IEEEproof}

Corollary \ref{CoroBackhaulInsights} shows a surprisingly simple dependence of backhaul configurations vs. overhead quality requirements: the lower bounds of $\bar{\mu}$ -- the average bit rate of backhaul servers -- are proportional to the overhead packet size $B$ and arrival rate $\eta$ while inversely proportional to the delay deadline $d$. Such lower bounds are expected to be tight, since they are based on the tight bounds in Remark \ref{ReLBGeneral}. As seen from the proof, the lower bound can be indeed achieved under: 1) deterministic overhead arrivals; and 2) equal backhaul rate allocation (stated in Remark \ref{ReOptiBackhaul}).

\subsection{Special Cases: Deterministic and Poisson Overhead Arrivals}
\begin{corollary} \label{CoroBackhaulAC}
\emph{For backhaul signaling under deterministic overhead arrivals, the overhead quality contour is}
\begin{align}
\mathcal{Q}_o = \left\{(\mathcal{T},B, d,p_e): p_e = 1- \mathcal{F}\left(\min(d,1/\eta), B, \{\mu_i\}|_{i=1}^N\right) \right\}.
\end{align}
\end{corollary}
\begin{IEEEproof}
Deterministic overhead arrival corresponds to the case of $\mathcal{T} \sim $ Gamma $\left(M,\frac{1}{M\eta}\right)$ with $M \rightarrow \infty$. Before we proceed to derive overhead rate and delay contours, we state two important results below.
\begin{align} \label{EquProofCoroBackhaulAC}
\lim_{M \rightarrow \infty} \left(\frac{M\eta}{M \eta + x}\right)^M & = \lim_{M \rightarrow \infty} \left(1-\frac{x/\eta}{M +x/\eta}\right)^M =e^{-x/\eta}. \\
\lim_{M \rightarrow \infty} \frac{\gamma(M,M x)}{\Gamma (M)} &= \lim_{M \rightarrow \infty} \frac{\int_{0}^{Mx} u^{M-1} e^{-u} \,\mathrm{d}u }{\int_{0}^\infty u^{M-1} e^{-u} \, \mathrm{d} u} = \mathbf{1}(x \geq 1).
\end{align}
Based on the equations immediately above, the outage probability $p_e$ is derived by letting $M \rightarrow \infty$ in Theorem \ref{ThmBackhual}.
\begin{align}
p_e &= \lim\limits_{M \rightarrow \infty}\sum\limits_{i=1}^N a_i \left[\left(1-\frac{\gamma(M,M\eta d)}{\Gamma(M)}\right)e^{-\frac{\mu_i d}{B}}+\frac{\gamma\left(M,M\eta d+\frac{\mu_i d}{B}\right)}{\Gamma(M)} \left(\frac{M\eta}{M\eta+\frac{\mu_i}{B}}\right)^M\right] \nonumber \\
&= \sum\limits_{i=1}^N a_i \left[(1-\mathbf{1}(\eta d \geq 1))e^{-\mu_i d/B}+ \mathbf{1}(\eta d \geq 1) e^{-\frac{\mu_i}{\eta B}}\right] \nonumber \\
&=
\begin{cases}
\sum\limits_{i=1}^N a_i e^{-\frac{\mu_i}{\eta B}} & d \geq 1/\eta \\
\sum\limits_{i=1}^N a_i e^{-\mu_i d /B} & d < 1/\eta \\
\end{cases}
\nonumber \\
&=1-\mathcal{F}\left(\min(d,1/\eta),B, \{\mu_i\}|_{i=1}^N\right)
\end{align}
\end{IEEEproof}

Under deterministic overhead arrivals, the lower bound $p_e^{lb,2} $ in (\ref{EquBackhaulepsilonlb2}) is simplified as
\begin{align} \label{EquBackhaulepsilonlb2AC}
p_e^{lb,2}= \lim_{M \rightarrow \infty} \sum\limits_{i=1}^N a_i \left(\frac{M\eta}{M\eta+\mu_i/B}\right)^M\stackrel{(a)}{=} \sum\limits_{i=1}^N a_i e^{-\frac{\mu_i}{\eta B}} = 1-\mathcal{F} (1/\eta,B, \{\mu_i\}|_{i=1}^N),
\end{align}
where (a) holds directly from (\ref{EquProofCoroBackhaulAC}). Combining the two lower bounds under deterministic overhead arrivals, i.e. (\ref{EquBackhaulepsilonlb1}) and (\ref{EquBackhaulepsilonlb2AC}), we have
\begin{align}
\max\left(p_e^{lb,1}, p_e^{lb,2}\right) = 1-\mathcal{F} (\min(d,1/\eta), B, \{\mu_i\}|_{i=1}^N).
\end{align}
It is important to note that the lower bound above is \emph{exactly} $p_e$ given in Corollary \ref{CoroBackhaulAC}.
\begin{remark} \label{ReLBAC}
\emph{Deterministic overhead arrivals minimize the outage probability, by achieving the lower bound given in (\ref{EquLBGeneral}).}
\end{remark}
The above remark implies that ignoring the randomness in the overhead process, previous models capture the lower bound on the overhead outage. Numerical results show that this lower bound is not tight.
\begin{corollary} \label{CoroBackhaulPC}
\emph{For backhaul signaling with Poisson overhead arrivals, the overhead quality contour is}
\begin{align}
\mathcal{Q}_o = &\left \{ (\mathcal{T},B,d,p_e): p_e = 1-\left(\prod\limits_{i=1}^N \frac{\mu_i}{\mu_i+\eta B}\right) \mathcal{F} (d, B, \{\mu_i+\eta B\})\right\}.
\end{align}
\end{corollary}
\begin{IEEEproof}
Poisson overhead arrivals correspond to the special case of $\mathcal{T} \sim $ Gamma $\left(M,\frac{1}{M\eta}\right)$ with $M=1$.
\begin{align}
p_e &\stackrel{(a)}{=} \sum\limits_{i=1}^N a_i \left[e^{-\eta d+\mu_i d/B} +\left(1-e^{-\eta d +\mu_i d/B}\right) \left(\frac{\eta}{\eta+\mu_i/B}\right)\right] \nonumber \\
&\stackrel{(b)}{=}1-\sum\limits_{i=1}^N a_i \left(1-e^{-\eta d +\mu_i d/B}\right) \left(1-\frac{\eta}{\eta+\mu_i/B} \right)\nonumber \\
&\stackrel{(c)}{=}1-\left(\prod\limits_{i=1}^N \frac{\mu_i}{\mu_i+\eta B}\right) \mathcal{F} (d,B, \{\mu_i+\eta B\}|_{i=1}^N).
\end{align}
The equality (a) comes from the fact that $\gamma(1,x)= 1-e^{-x}$ and $\Gamma (1)=1$, while equality (b) holds from Property \ref{PropertySum} (letting $x=0$). See the proof of Property \ref{PropertySum} for equality (c).
\end{IEEEproof}

For a given sum of service rates $\sum_{i=1}^N \mu_i=C$, the delay CDF $\mathcal{F} (d,B,\{\mu_i\}|_{i=1}^N)$ is maximized for any $d$ and $B$ \emph{iff} all service rates are equal, i.e. $\mu_i=\bar{\mu}=C/N$ ($1 \leq i \leq N$). Therefore equal rate allocation among backhaul servers minimizes the overhead outage under deterministic arrivals in Corollary \ref{CoroBackhaulAC}. It is also the optimal choice for Poisson overhead arrivals because it simultaneously maximizes $\prod_{i=1}^N \frac{\mu_i}{\mu_i+\eta B}$ and $\mathcal{F} (d, B, \{\mu_i+\eta B\}|_{i=1}^N)$ in Corollary \ref{CoroBackhaulPC}. Such a conclusion in fact holds under general overhead arrivals. The maximized CDF implies that the delay $\mathcal{D}$ is stochastically minimized. The outage probability $p_e = 1-\mathbb{P} (\mathcal{D} \leq \mathcal{T}, \mathcal{D} \leq d)$ is therefore minimized, independent on the overhead arrival process.

\begin{remark} \label{ReOptiBackhaul}
\emph{For a given sum of service rates, equal rate allocation among backhaul servers minimizes the overhead outage, independent on overhead arrival process.}
\end{remark}
Remark \ref{ReLBAC} and \ref{ReOptiBackhaul} together imply that the overhead outage is minimized when overhead arrivals are deterministic \emph{and} backhaul servers have the same overhead processing rate $\bar{\mu}$.

\section{Overhead Quality Contour in Wireless Signaling}
Dedicated wireless links (e.g. out-of-band GSM or to-be-defined overhead channels in LTE-A) are also leveraged by coordination techniques to share overhead (e.g. CSI feedback). Since the radio environment in HCNs is very different from $1$-tier macrocell case, the wireless overhead channels present new characteristics such as SINR distributions. In this section, we quantify the \emph{overhead quality contour} for wireless signaling in HCNs. As seen from Section II, the wireless link delay is contingent on which tier $\mathrm{BS}_n$ belongs to. In this section, denote the tier index of $\mathrm{BS}_n$ as $k$ ($1 \leq k \leq K$).

\subsection{General Case and Main Results}
\begin{theorem} \label{ThmWireless}
\emph{For wireless overhead signaling between $\mathrm{BS}_n$ and $\mathrm{BS}_0$ with interarrival time $\mathcal{T} \sim $ \emph{Gamma} $\left(M,\frac{1}{M\eta}\right)$, the overhead quality contour is}
\begin{align} \label{EquThmWireless}
\mathcal{Q}_o = \left\{(\mathcal{T},B, d, p_e): p_e = \left[1-\frac{\gamma(M,M\eta d)}{\Gamma(M)}\right] q_k\{\beta (d)\} + \int_{0}^d q_k\{\beta (x)\} \frac{(M\eta x)^{M} e^{-M\eta x} }{x \Gamma(M)}\mathrm{d}x  \right\},
\end{align}
\emph{where $\beta(x)=\exp\left(\frac{B\ln 2}{Wx}\right)-1$ is the required SIR for overhead deadline $x$ and the subscript k is the tier index of $\mathrm{BS}_n$.}
\end{theorem}
\begin{IEEEproof}
See Appendix \ref{ProofThmWireless}.
\end{IEEEproof}

Theorem \ref{ThmWireless} quantifies the possible pairs of $(\mathcal{T},B,d,p_e)$ for arbitrary wireless overhead channel setups. However, for the same reason stated below Theorem \ref{ThmBackhual}, we derive simpler bounds on $\mathcal{Q}_o$ in (\ref{EquThmWireless}).
\begin{remark} \label{ReLBWireless}
\emph{The overhead outage in Theorem \ref{ThmWireless} can be bounded as}
\begin{align} \label{EquWirelessBounds}
\max (q_k\{\beta(d)\},\mathbb{E}[q_k\{\beta(\mathcal{T})\}]) \leq p_e \leq \left[1-\frac{\gamma(M,M\eta d)}{\Gamma(M)}\right] q_k\{\beta (d)\} + \frac{\gamma(M,M\eta d)}{\Gamma(M)}\,.
\end{align}
\end{remark}
Using the same argument in Section III, the lower bound on overhead outage can be achieved by letting $\eta \rightarrow 0$ or $d \rightarrow \infty$. The upper bound on $p_e$ can be found based on the fact that $q_k\{\cdot\} \leq 1$. By restricting several overhead parameters in (\ref{EquWirelessBounds}) as required by coordination techniques, the bounds determine their feasible overhead sets in an easier way than Theorem \ref{ThmWireless}.

Remark \ref{ReLBWireless} shows the clear dependence between overhead outage $p_e$ and the distribution of SINR ($q_k\{\cdot\}$ function) and $\mathcal{T}$ ($\gamma(\cdot)$ function). Therefore with appropriate models on  SINR and overhead interarrival time $\mathcal{T}$, the \emph{overhead quality contour} in Theorem \ref{ThmWireless} provides more accurate insights than previous works on feasible overhead parameters in HCNs.
\begin{corollary} \label{CoroWirelessOpti}
\emph{When $BSs$ of all tiers have the same path loss exponent $\alpha$, for a given overhead requirement $(\mathcal{T},B,d,p_e)$ from coordination techniques, the bandwidth $W$ of wireless overhead channel must satisfy following inequalities}
\begin{align} \label{EquWLB}
W \geq \frac{B}{d \log \left(1+\frac{\zeta(\alpha)}{(1-p_e)^{\alpha/2}}\right)}\,,
\end{align}
\emph{where} $\zeta(\alpha)= \left(\frac{2\pi}{\alpha}\csc{\frac{2\pi}{\alpha}}\right)^{-\frac{\alpha}{2}}$.
\end{corollary}
\begin{IEEEproof}
See Appendix \ref{ProofCoroWirelessOpti}.
\end{IEEEproof}

It is generally hard to provide design guidelines for wireless overhead channel (e.g. the bandwidth $W$) directly from the \emph{overhead quality contour} or even its bounds. This is mainly because of the complicated expression of $q_k\{\cdot\}$ as in Lemma \ref{LemmaSIR}. In Corollary \ref{CoroWirelessOpti} we discuss it in a special case where $q_k\{\cdot\}$ can be simplified. The discussion under general form of $q_k\{\cdot\}$ will be left to numerical results in Section V.
\subsection{Special Cases: Deterministic and Poisson Overhead Arrivals}
\begin{corollary} \label{CoroWirelessAC}
\emph{For wireless signaling with deterministic overhead arrivals, the overhead quality contour is}
\begin{align}
\mathcal{Q}_o = \left\{(\mathcal{T},B,d,p_e): p_e = q_k\{\beta\left(\min(d,1/\eta)\right)\}\right\}.
\end{align}
\end{corollary}
\begin{IEEEproof}
Based on the proof of Theorem \ref{ThmWireless}, overhead outage under deterministic overhead arrival is
\begin{align}
p_e &= 1-\lim_{M\rightarrow \infty}\int_{0}^d \left[1-\frac{\gamma(M,M\eta x)}{\Gamma(M)}\right] \mathrm{d} \mathbb{P} (\mathcal{D} \leq x) \nonumber \\
&= 1-\int_{0}^d [1-\mathbf{1}(\eta x \geq 1)] \mathrm{d} \mathbb{P} (\mathcal{D} \leq x) \nonumber \\
&=1- \mathbb{P} (\mathcal{D} \leq \min(d,1/\eta)) \nonumber \\
&= q_k \{\beta (\min(d,1/\eta))\}.
\end{align}
\end{IEEEproof}
\begin{corollary} \label{CoroWirelessPC}
\emph{For wireless signaling with Poisson overhead arrivals, the overhead quality contour is}
\begin{align}
\mathcal{Q}_o = \left\{(\mathcal{T},B,d,p_e): p_e = e^{-\eta d} q_k\{\beta(d)\} + \int_{0}^d q_k \{\beta(x)\} \eta e^{-\eta x} \mathrm{d} x\right\}.
\end{align}
\end{corollary}
\begin{IEEEproof}
The proof follows simply by replacing the general expression $\gamma(M,x)$ with $\gamma(1,x) = 1-e^{-x}$.
\end{IEEEproof}

The results on $\emph{overhead quality contour}$ in Theorem \ref{ThmWireless} are greatly simplified in these special cases. Under deterministic arrivals, the lower bound on $p_e$ in Remark \ref{ReLBWireless} reduces to
\begin{align}
\max (q_k\{\beta(d)\},\mathbb{E}[q_k\{\beta(\mathcal{T})\}]) = \max (q_k\{\beta(d)\},q_k\{\beta(1/\eta)\}) = q_k\{\beta\left(\min(d,1/\eta)\right)\}.
\end{align}
It is seen that, similar to backhaul signaling, deterministic arrivals are also optimal in wireless signaling. In other words, ignoring natural randomness in overhead arrivals leads to underestimation of wireless overhead delay and outage, which is non-trivial as shown through numerical results below.

\section{Numerical Results and Discussion}
In this section, we consider a $3$-tier heterogeneous network as shown in Fig. \ref{PicSystemModel}, comprising for example macro (tier $1$), pico (tier $2$) and femto (tier $3$) BSs. Notation and system parameters are given in Table \ref{table1}. Suppose $\mathrm{BS}_0$ is a pico BS. According to the tier index of $\mathrm{BS}_n$, the \emph{overhead quality contour} is investigated in the following three scenarios.

\textbf{Scenario I: $\mathbf{BS_n}$ belongs to $\mathbf{1^{st}}$ tier.} The backhaul path between pico and macro BSs includes backhaul servers from the core network and the picocell aggregator, i.e. $N=N_{CN}+1$.

\textbf{Scenario II: $ \mathbf{BS_n}$ belongs to $\mathbf{2^{nd}}$ tier.} Since nearby pico BSs are often clustered by sharing the same backhaul aggregator \cite{TROPOS07}, the number of backhaul servers between $\mathrm{BS}_0$ and its neighbor $\mathrm{BS}_n$ is $N=1$.

\textbf{Scenario III: $\mathbf{BS_n}$ belongs to $\mathbf{3^{rd}}$ tier.} The backhaul servers between pico and femto BSs consist of the picocell aggregator, the femtocell gateway, and those from the core network and femtocell's IP network, i.e. $N=2+N_{CN}+N_{IP}$.

In all three scenarios, we assume all backhaul servers have the same rate $\bar{\mu}$ for overhead packets, which is optimal per Remark \ref{ReOptiBackhaul}.
\subsection{Overhead Quality Contour in Backhaul Signaling}
\begin{figure}[hp]
\centerline{
\includegraphics[width=4in]{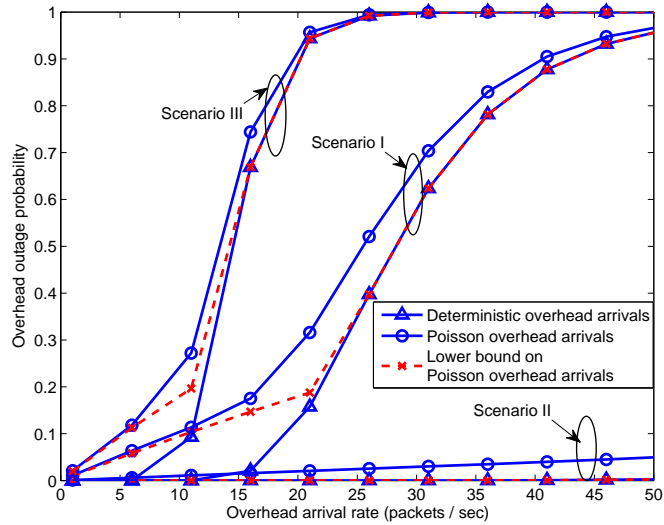}}
\caption{Overhead outage $p_e$ vs. overhead arrival rate $\eta$ in all three scenarios. The delay requirement $d$ is $0.3 \,\mathbb{E} [\mathcal{T}] = 0.3/\eta$, i.e. overhead signaling is allowed to occupy $30\%$ time slots. The overhead service rate $\frac{\bar{\mu}}{B}=1000$ packets/sec.} \label{PicBhRate2}
\end{figure}

\begin{figure}[hp]
\centerline{
\includegraphics[width=4in]{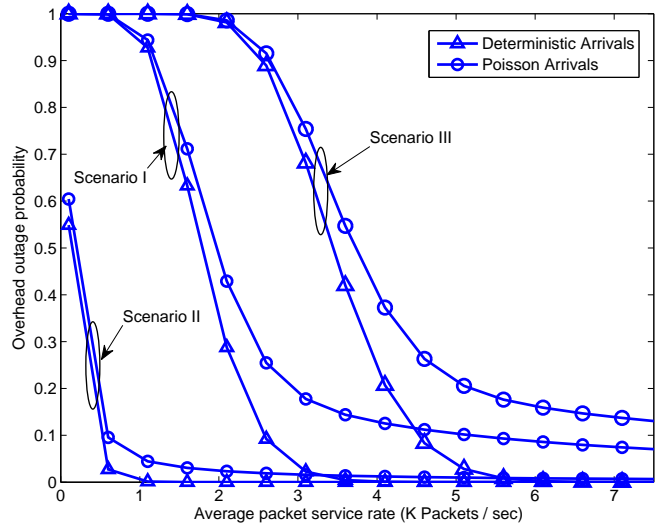}}
\caption{Overhead outage $p_e$ vs. average packet service rate $\bar{\mu}/B$ in the three scenarios. The overhead rate $\eta=50$ packets/sec, i.e. an overhead on average has lifetime $\mathbb{E}(\mathcal{T}) = 1/\eta=20$ ms. The overhead delay requirement $d$ is $0.3\, \mathbb{E} [\mathcal{T}] = 6$ ms.} \label{PicBhMu}
\end{figure}

\textbf{$\mathcal{Q}_o$ vs. Backhaul Configurations.} The \emph{overhead quality contour} in various backhaul configurations (i.e. number of backhaul servers $N$ and their rate $\bar{\mu}$) is shown in Fig. \ref{PicBhRate2} and \ref{PicBhMu}. Obviously, the overhead outage decreases as the number of servers decreases and/or their rate $\bar{\mu}$ increases. However two important observations are worth noting: 1) reducing the number of backhaul servers is critically important since the outage in scenarios II ($N=1$) is way below the other two scenarios ($N >10$); 2) it is difficult to ensure very small outage (e.g. $p_e \leq 0.1$) purely by increasing backhaul servers' rate $\bar{\mu}$, since the outage curve in Fig. \ref{PicBhMu} is almost flat in the region of $p_e \leq 0.1$. Under this circumstance, our conjecture is that appropriate retransmission schemes and certain level of coding should also be deployed for further outage reduction.

\textbf{Insights on Backhaul Deployments.} According to the specific overhead requirements, the minimum rate of backhaul servers is derived in Corollary \ref{CoroBackhaulInsights} based on the lower bound in Remark \ref{ReLBGeneral}. This bound is achieved under deterministic arrivals (as stated in Remark \ref{ReLBAC}) but suspected to be loose under Poisson arrivals -- the opposite extreme of deterministic. However Fig. \ref{PicBhRate2} shows that it is fairly tight even for Poisson arrivals, especially in small outage region (i.e. $p_e \leq 0.1$) of practical interest. Therefore, the results in Corollary \ref{CoroBackhaulInsights} provide accurate guidelines on the deployment of backhaul overhead channels in HCNs.

\textbf{Comparison with Previous Models.} Fig. \ref{PicBhRate2} also shows the appreciable difference in overhead outage between Poisson and deterministic arrivals. For example, with the same overhead rate of $10$ packets/sec in scenario III, deterministic arrivals incur $0.1$ outage (usually an acceptable packet error percentage) while Poisson arrivals incur $0.3$ outage (generally unacceptable). In short, the randomness in overhead arrivals is an important factor for overhead signaling characterization but missed from previous works.

\begin{figure}[hp]
\centerline{
\includegraphics[width=4in]{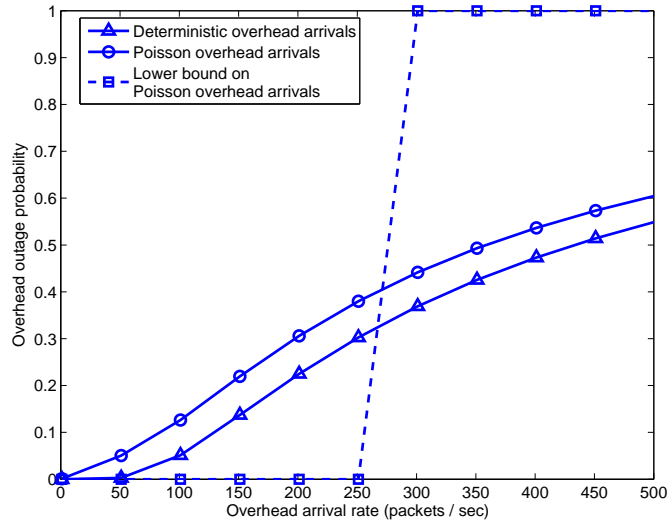}}
\caption{Overhead outage $p_e$ vs. overhead arrival rate $\eta$ in scenario II. The delay requirement $d$ is $0.3 \,\mathbb{E} [\mathcal{T}] = 0.3/\eta$, i.e. overhead signaling is allowed to occupy $30\%$ time slots. The overhead service rate $\frac{\bar{\mu}}{B}=1000$ packets/sec and the number of backhaul servers $N=1$, which together translate to a mean delay of $\mathbb{E} (\mathcal{D}) = 1$ ms. Previous simplified models assume constant overhead delay $\mathcal{D}=\mathbb{E}(\mathcal{D}) = 1$ ms and constant overhead arrivals $\mathcal{T} = \mathbb{E}(\mathcal{T})=1/\eta$. } \label{PicBhRate}
\end{figure}

Fig. \ref{PicBhRate} shows the more comprehensive comparison of our results with previous simplified models in scenario II. It is seen that previous simplified models, ignoring the randomness in overhead arrivals \emph{and} backhaul delay, are highly inaccurate even though their underlying assumption of low-latency backhaul interface is satisfied in scenario II (mean delay is $1$ ms). Under an outage requirement of, for example, $p_e \leq 0.1$, they predict that backhaul channel can support up to $250$ packets/sec, which in fact is between $75$ (Poisson arrivals) and $125$ packets/sec (deterministic arrivals).

\subsection{Overhead Quality Contour in Wireless Signaling}
\begin{figure}[hp]
\centerline{
\includegraphics[width=4in]{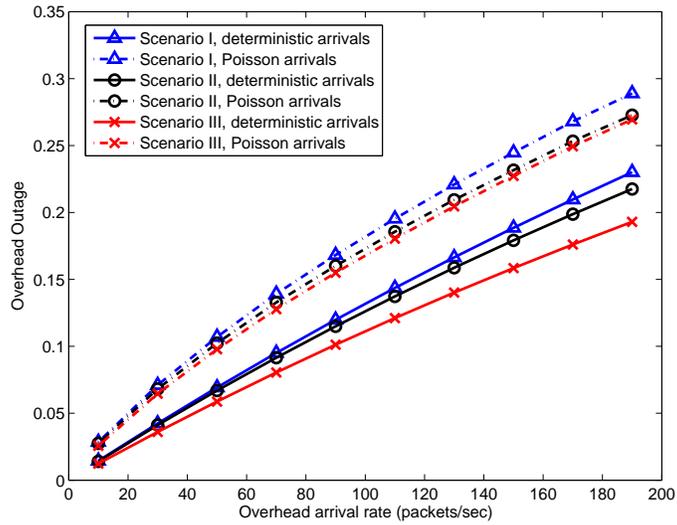}}
\caption{Overhead outage $p_e$ vs. overhead arrival rate $\eta$ for wireless signaling. The delay requirement $d$ is $0.3 \,\mathbb{E} [\mathcal{T}] = 0.3/\eta$, i.e. overhead signaling is allowed to occupy $30\%$ time slots. The overhead channel bandwidth is $50$ KHz.} \label{PicWiRate}
\end{figure}

\textbf{$\mathcal{Q}_o$ vs. Overhead Channel Configurations.} Fig. \ref{PicWiRate} shows the overhead outage $p_e$ in three scenarios, i.e. different types of $\mathrm{BS}_n$. For a given arrival process, the outage curves of different scenarios are very close to each other. It is somewhat counter-intuitive since different types of $\mathrm{BS}_n$ have different powers, path loss exponents and wall penetrations. The underlying reason comes from the fact that $\mathrm{BS}_n$ has the strongest received power at $\mathrm{BS}_0$. If it has low transmitting power, large path loss exponent and wall penetration, it must be close to $\mathrm{BS}_0$. Thus the statistics of overhead channel between $\mathrm{BS}_n$ and $\mathrm{BS}_0$ are roughly independent of the types of $\mathrm{BS}_n$, and so is the \emph{overhead quality contour}.

\begin{figure}[hp]
\centerline{
\includegraphics[width=4in]{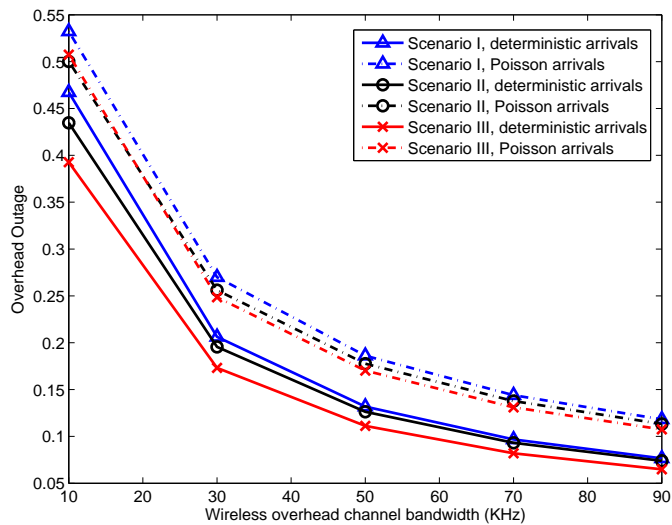}}
\caption{Overhead outage $p_e$ vs. wireless overhead channel bandwidth $W$. The overhead rate $\eta=100$ packets/sec, and the delay requirement $d$ is $0.3 \,\mathbb{E} [\mathcal{T}] = 0.3/\eta$, i.e. overhead signaling is allowed to occupy $30\%$ time slots.} \label{PicWiW}
\end{figure}

Fig. \ref{PicWiW} illustrates the outage $p_e$ vs. wireless overhead channel bandwidth $W$. The observation here is similar to Fig. \ref{PicBhMu}: increasing bandwidth can easily reduce outage to about $0.1$ but is not a cost-effective way for further outage reduction. Therefore, retransmission schemes, coding and diversity techniques will be useful in this situation.

\begin{figure}[hp]
\centerline{
\includegraphics[width=4in]{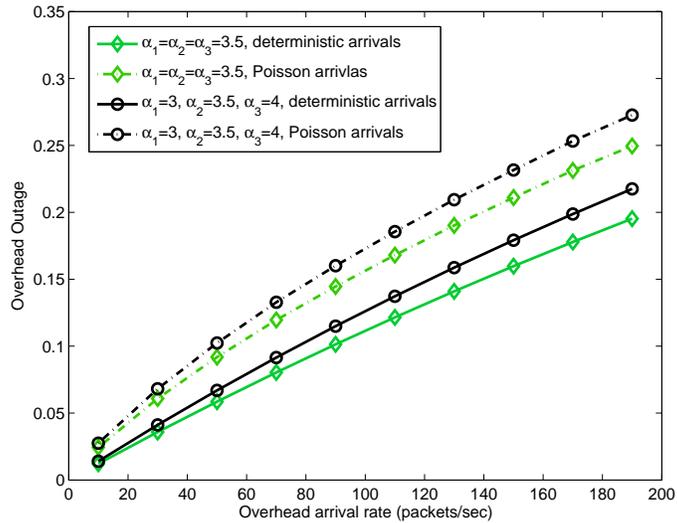}}
\caption{Overhead outage $p_e$ under equal vs. different path loss exponents (as listed in Table \ref{table1}). $\mathrm{BS}_n$ is assumed to be pico BS, i.e. it belongs to the second tier. The delay requirement $d$ is $0.3 \,\mathbb{E} [\mathcal{T}] = 0.3/\eta$.} \label{PicWiAlpha}
\end{figure}

\textbf{Insights on Wireless Overhead Channel Deployment.} Assuming equal path loss exponents, Corollary \ref{CoroWirelessOpti} quantifies the minimum bandwidth $W$ of wireless overhead channel under given overhead requirements. Fig. \ref{PicWiAlpha} shows that such a simplified assumption is surprisingly reasonable: the outage $p_e$ is almost the same (with difference less than $0.02$) under equal or different path loss exponents. Thus, the insights on overhead channel deployment in Corollary \ref{CoroWirelessOpti} are predicted to be accurate as well.

\textbf{Comparison with Previous Models.} Two key differences from previous models contribute to our more accurate characterization of the overhead signaling in HCNs: 1) the consideration of overhead arrival dynamics, because Fig. \ref{PicWiRate} shows that Poisson overhead arrivals incur higher outage than deterministic arrivals (no randomness in $\mathcal{T}$ as assumed in previous models) by $0.05$ to $0.1$; 2) the appropriate spatial model on BS locations in HCNs, which is fundamental to spatial interference statistics and overhead channel SINR distribution $q_k\{\cdot\}$. The comparison of spatial models (our PPP based model vs. previous assumed grid model) is extensively discussed in \cite{FleStoSim97,AndBacGan10,DhiGanBacAnd11,JoSanXiaAnd11}.

\subsection{The Optimal Overhead Signaling Method}
\begin{figure}[htp]
\centerline{\subfigure[Deterministic overhead arrivals]{\includegraphics[width=3.5in]{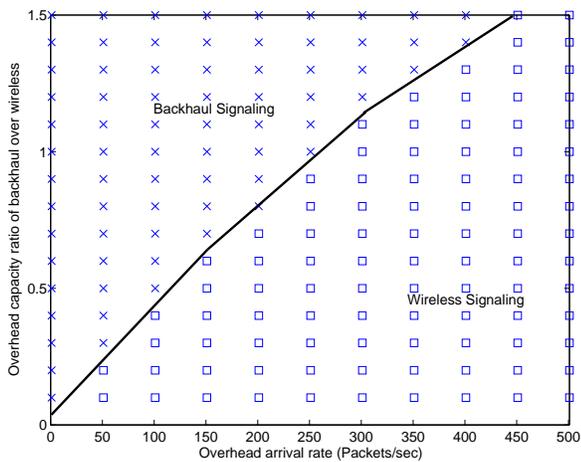} }
\hfil \subfigure[Poisson overhead arrivals]{\includegraphics[width=3.5in]{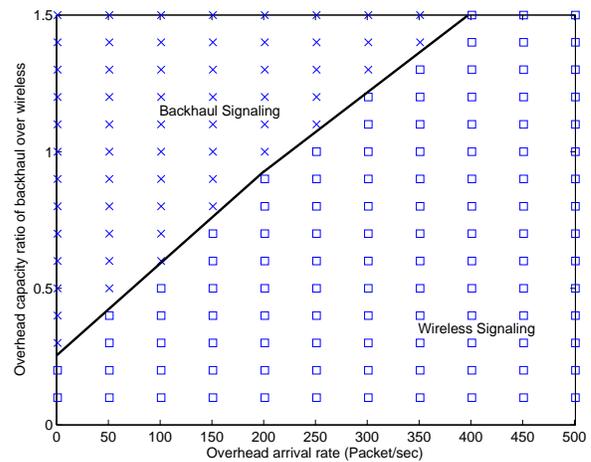} }}
\caption{Optimal overhead channel choice in Scenario I under deterministic and Poisson overhead arrivals. The wireless overhead channel bandwidth is $50$ KHz and its overhead capacity $\mathcal{R}_{Wi}= 1/\mathbb{E} [\mathcal{D}] \doteq 1000$ packets/sec. The delay requirement $d$ is $0.3 \,\mathbb{E} [\mathcal{T}] = 0.3/\eta$. The mark ``$\Box$'' means wireless signaling is preferred with lower outage, while ``$\times$'' means backhaul signaling is preferred.} \label{Picchoice}
\end{figure}

Numerical results show that in all three scenarios, the optimal choices between backhaul vs. wireless signaling are determined by two important measures: 1) the overhead arrival rate $\eta$; and 2) the overhead capacity of backhaul and wireless overhead channel, which is defined as the inverse of average overhead delay in (\ref{EquBackhaulLatency}) and (\ref{EquWirelessLatency})
\begin{align}
\mathcal{R} = \frac{1}{\mathbb{E}[\mathcal{D}]} =
\begin{cases}
\frac{\bar{\mu}}{N B} \stackrel{\triangle}{=} \mathcal{R}_{Ba} & \mbox{backhaul overhead channel} \\
\frac{W}{B}\mathbb{E} [\log(1+\mbox{SIR})] \stackrel{\triangle}{=} \mathcal{R}_{Wi} & \mbox{wireless overhead channel}\\
\end{cases}
\end{align}
Fig. \ref{Picchoice} depicts the optimal choice in Scenario I under deterministic and Poisson arrivals. In general, the backhaul channel is preferred for slow overhead traffic. For example, it can serve overhead traffic of $50$ packets/sec with only $30\%  \sim 50\% $ of the overhead capacity of wireless channel. On the other hand, the wireless channel is more preferred for fast overhead sharing. Comparing Fig. \ref{Picchoice} (a) and (b), it is seen that as the randomness of overhead arrivals increases, wireless signaling becomes more preferable.

\section{Conclusion}
This paper has presented a new framework to quantify the feasible set of inter-cell overhead delay, rate and outage as a function of plausible HCN deployments. This framework allows a more realistic but analytically tractable assessment on inter-cell coordination in HCNs by quantifying the inherent impact of the overhead signaling. It also provides design guidelines on HCN overhead channels (backhaul and wireless) to accommodate specific coordination techniques. Future extensions to this approach can include sophisticated overhead retransmission schemes or overhead signaling between multiple (more than two) cells.

\section*{Acknowledgement}
The authors gratefully acknowledge Dr. Amitava Ghosh and Dr. Bishwarup Mondal of Motorola Solutions (Now Nokia Siemens) for their valuable technical inputs and feedback regarding this paper.

\appendix
\subsection{Proof of Property \ref{PropertySum}} \label{ProofPropertySum}
For $x=0$, Property \ref{PropertySum} obviously holds, since
\begin{align} \label{EquProofProperty}
\sum_{i=1}^N a_i \frac{\mu_i}{\mu_i+x} = \sum_{i=1}^N a_i = \sum_{i=1}^N a_i \lim_{d\rightarrow \infty} (1-e^{-\mu_i d/B})=\mathcal{F} (\infty, B, \{\mu_i\}|_{i=1}^N)=1
\end{align}
For $x>0$, we have
\begin{align}
\sum_{i=1}^N a_i \frac{\mu_i}{\mu_i+x} &= \sum_{i=1}^N\prod\limits_{j \neq i} \frac{\mu_j}{\mu_j-\mu_i} \frac{\mu_i}{\mu_i+x}
= \left(\prod\limits_{i=1}^N \frac{\mu_i}{\mu_i+x}\right) \sum\limits_{i=1}^N \prod\limits_{j\neq i}\frac{\mu_j+x}{\mu_j-\mu_i} \stackrel{(a)}{=} \left(\prod\limits_{i=1}^N \frac{\mu_i}{\mu_i+x}\right)
\end{align}
Note that $\{\prod\limits_{j\neq i}\frac{\mu_j+x}{\mu_j-\mu_i}\}|_{i=1}^N$ is indeed the coefficient $\{a_i\}|_{i=1}^N$ in $\mathcal{F} (d, B, \{\mu_i+x\}|_{i=1}^N)$. Thus the equality (a) holds from (\ref{EquProofProperty}).
\subsection{Proof of Theorem \ref{ThmBackhual}} \label{ProofThmBackhaul}
According to its definition, the successful overhead will not be dropped by the backhaul servers. Therefore its delay is the sum latencies from all the backhaul servers as in (\ref{EquBackhaulLatency}). With the delay CDF given in (\ref{EqubackhaulCDF}), the overhead outage is derived as
\begin{align}
p_e &= 1-\int_{0}^{d} \mathbb{P} (\mathcal{T} \geq x) \;\mathrm{d} \mathcal{F}(x, B, \{\mu_i\}|_{i=1}^N)\nonumber \\
&=1- \int_{0}^{d} \left[1-\frac{\gamma(M,M\eta x)}{\Gamma(M)}\right] \mathrm{d} \mathcal{F}(x, B, \{\mu_i\}|_{i=1}^N)\nonumber \\
&=1-\mathcal{F}(d, B, \{\mu_i\}|_{i=1}^N) +\sum\limits_{i=1}^N a_i\int_{0}^d \left(\int_{0}^{M\eta x} \frac{u^{M-1} e^{-u}}{\Gamma (M)} \mathrm{d} u\right) \; \frac{\mu_i e^{-\mu_i x/B}}{B} \mathrm{d}x \nonumber \\
&=1-\sum\limits_{i=1}^N a_i \left[1-e^{-\mu_i d/B}-\frac{\gamma(M,M\eta d+\mu_i d/B)}{\Gamma(M)} \left(\frac{M\eta}{M\eta+\mu_i}\right)^M+\frac{\gamma(M,M\eta d)}{\Gamma(M)}e^{-\mu_i d/B}\right]\nonumber \\
&\stackrel{(a)}{=}\sum\limits_{i=1}^N a_i \left[\left(1-\frac{\gamma(M,M\eta d)}{\Gamma(M)}\right)e^{-\frac{\mu_i d}{B}}+\frac{\gamma\left(M,M\eta d+\frac{\mu_i d}{B}\right)}{\Gamma(M)} \left(\frac{M\eta}{M\eta+\frac{\mu_i}{B}}\right)^M\right]
\end{align}
The equality (a) comes from the fact that $\sum_{i=1}^N a_i=1$ (Property \ref{PropertySum} by letting $x=0$).
\subsection{Proof of Corollary \ref{CoroBackhaulInsights}} \label{ProofCoroBackhaulInsights}
As seen in Remark \ref{ReOptiBackhaul}, equal rate allocation minimizes the overhead outage for a given sum rate. Under this backhaul setup, the CDF of delay $\mathcal{D}$ as in (\ref{EquBackhaulLatency}) is gamma distributed with CDF given as
\begin{align}
\mathbb{P} (\mathcal{D} \leq d) = \frac{\gamma (N,B d/\bar{\mu})}{\Gamma(N)},
\end{align}
where $\bar{\mu}=\frac{\sum_{i=1}^N \mu_i}{N}$. Based on the proof of Theorem \ref{ThmBackhual}, we have
\begin{align}
p_e & \geq1- \int_{0}^{d} \left[1-\frac{\gamma(M,M\eta x)}{\Gamma(M)}\right] x^{N-1} \left(\frac{\bar{\mu}}{B}\right)^N \frac{e^{-\bar{\mu}x/B}}{\Gamma(N)} \mathrm{d}x\nonumber \\
&=1-\frac{\bar{\mu}^N}{B^N (N-1)!}\sum_{k=0}^{M-1}\int_{0}^d \frac{(M\eta x)^k x^{N-1} }{k!} e^{-M\eta x+\bar{\mu}x/B} \mathrm{d}x \nonumber \\
&\stackrel{(a)}{\geq} \max\left(1-\frac{\gamma(N,\bar{\mu}d/B)}{(N-1)!},\; 1-\frac{\bar{\mu}^N}{B^N (N-1)!}\sum_{k=0}^{M-1} \frac{(M\eta)^k (k+N-1)!}{k! (M\eta+\bar{\mu}/B)^{k+N}}\right)\nonumber \\
&=\max \left(1-\frac{\gamma(N,\bar{\mu}d/B)}{(N-1)!}, \;1-\sum_{k=0}^{M-1} \binom{k+N-1}{N-1} p^k (1-p)^{N}\right),
\end{align}
where $p=\frac{M\eta}{M\eta+\bar{\mu}/B}$. Based on the argument in Remark \ref{ReLBGeneral}, inequality (a) follows by letting $\eta=0$ or $d =\infty$. For a given overhead requirement $(\mathcal{T},B,\mathcal{D},p_e)$, the value of $\bar{\mu}$ and $N$ must satisfy
\begin{align}
p_e & \geq 1-\frac{\gamma(N,\bar{\mu}d/B)}{(N-1)!} \Rightarrow \bar{\mu}  \geq \frac{B}{d}  \,\gamma^{-1} \left((1-p_e) (N-1)!, N\right) \\
p_e & \geq 1-\sum_{k=0}^{M-1} \binom{k+N-1}{N-1} p^k (1-p)^{N} \nonumber \\
& \stackrel{(b)}{\geq} 1- \binom{N+M-1}{N} (1-p)^{N} \stackrel{(c)}{\Rightarrow}
\bar{\mu}  \geq \frac{M \eta B\sqrt[N]{1-p_e}} {\sqrt[N]{\binom{M+N-1}{N}}-\sqrt[N]{1-p_e}}
\end{align}
Inequality (b) follows from $p^k \leq 1$ and (c) holds by substituting back for $p$.
\subsection{Proof of Theorem \ref{ThmWireless}} \label{ProofThmWireless}
The outage probability $p_e$ in wireless signaling is
\begin{align}
p_e &= 1- \mathbb{P} (\mathcal{D} \leq d, \mathcal{D} \leq \mathcal{T}) \nonumber \\
&= 1- \int_{0}^d \mathbb{P} (\mathcal{T} \geq x) \mathrm{d} \mathbb{P} (\mathcal{D} \leq x) \nonumber \\
&=1-\int_{0}^d \left[1-\frac{\gamma(M,M\eta x)}{\Gamma(M)}\right] \mathrm{d} \mathbb{P} (\mathcal{D} \leq x) \nonumber \\
&=\mathbb{P} (\mathcal{D} > d) + \frac{\gamma(M,M\eta d)}{\Gamma(M)}\mathbb{P}(\mathcal{D} \leq d)
- \int_0^d \mathbb{P}(\mathcal{D} \leq x) \frac{1}{\Gamma(M)}\mathrm{d} \gamma(M, M\eta x)
\end{align}
As $\mathrm{BS}_n$ belongs to the $k^{th}$ tier, the wireless overhead delay is characterized as
\begin{align}
\mathbb{P} (\mathcal{D} \leq x)  = \mathbb{P} \left(\frac{B}{ W \log (1+ \mbox{SIR})} \leq x \right)
&=1-q_k (\beta(x))
\end{align}
where $\beta(x) = \exp\left(\frac{B\ln 2}{W x}\right)-1$. The outage probability $p_e$ then follows.

\subsection{Proof of Corollary \ref{CoroWirelessOpti}} \label{ProofCoroWirelessOpti}
As shown in Corollary 2 in \cite{JoSanXiaAnd11}, the CDF $q_k\{\beta(d)\}$ is simplified under equal path loss exponents
\begin{align} \label{EquProofCoroOpti2}
q_k\{
\beta(d)\}=1-\frac{1}{1+\mathcal{Z}(\beta(d),\alpha)},
\end{align}
where the function $\mathcal{Z} (\beta(d),\alpha)$ is
\begin{align} \label{EquProofCoroOpti3}
\mathcal{Z}(\beta(d),\alpha) &=[\beta(d)]^{\frac{2}{\alpha}}\int_{[\beta(d)]^{-\frac{2}{\alpha}}}^{\infty} \frac{1}{1+u^{\frac{\alpha}{2}}} \mathrm{d} u \nonumber \\
&= [\beta(d)]^{\frac{2}{\alpha}}\int_{0}^{\infty} \frac{1}{1+u^{\frac{\alpha}{2}}} \mathrm{d} u -[\beta(d)]^{\frac{2}{\alpha}} \int_{0}^{[\beta(d)]^{-\frac{2}{\alpha}}} \frac{1}{1+u^{\frac{\alpha}{2}}} \mathrm{d} u \nonumber \\
&\geq [\beta(d)]^{\frac{2}{\alpha}}\frac{2\pi}{\alpha}\csc{\frac{2\pi}{\alpha}}-1.
\end{align}
Using the bound immediately above in the lower bound of $p_e$, (\ref{EquWLB}) follows.

\bibliographystyle{IEEEtran}
\bibliography{CoordMethod}

\begin{thebibliography}{10}
\providecommand{\url}[1]{#1}
\csname url@samestyle\endcsname
\providecommand{\newblock}{\relax}
\providecommand{\bibinfo}[2]{#2}
\providecommand{\BIBentrySTDinterwordspacing}{\spaceskip=0pt\relax}
\providecommand{\BIBentryALTinterwordstretchfactor}{4}
\providecommand{\BIBentryALTinterwordspacing}{\spaceskip=\fontdimen2\font plus
\BIBentryALTinterwordstretchfactor\fontdimen3\font minus
  \fontdimen4\font\relax}
\providecommand{\BIBforeignlanguage}[2]{{%
\expandafter\ifx\csname l@#1\endcsname\relax
\typeout{** WARNING: IEEEtran.bst: No hyphenation pattern has been}%
\typeout{** loaded for the language `#1'. Using the pattern for}%
\typeout{** the default language instead.}%
\else
\language=\csname l@#1\endcsname
\fi
#2}}
\providecommand{\BIBdecl}{\relax}
\BIBdecl

\bibitem{VC08CommMag}
V.~Chandrasekhar, J.~G. Andrews, and A.~Gatherer, ``Femtocell networks: A
  survey,'' \emph{IEEE Commun. Mag.}, vol.~46, no.~9, pp. 59--67, September
  2008.

\bibitem{Qcomm10}
A.~Khandekar, N.~Bhushan, J.~Tingfang, and V.~Vanghi, ``{LTE} advanced:
  Heterogeneous networks,'' in \emph{European Wireless Conference}, June 2010,
  pp. 978 -- 982.

\bibitem{TROPOS07}
``Picocell mesh: Bringing low-cost coverage, capacity and symmetry to mobile
  {WiMAX},'' White Paper, Tropos Network, March 2007.

\bibitem{ZhaAnd08}
J.~Zhang and J.~G. Andrews, ``Distributed antenna systems with randomness,''
  \emph{IEEE Transactions on Wireless Communications}, vol.~7, no.~9, pp.
  3636--3646, September 2008.

\bibitem{Gesbertetal10}
D.~Gesbert, S.~Hanly, H.~Huang, S.~S. Shitz, O.~Simeone, and W.~Yu,
  ``Multi-cell {MIMO} cooperative networks: A new look at interference,''
  \emph{IEEE J. Select. Areas Commun.}, vol.~28, no.~9, pp. 1380 -- 1408,
  December 2010.

\bibitem{SR09CSI}
S.~Ramprashad and G.~Caire, ``Cellular vs. network {MIMO}: A comparison
  including the channel state information overhead,'' in \emph{Proc. of the
  IEEE Int. Symp. on Personal Indoor and Mobile Radio Comm.}, September 2009.

\bibitem{SR09Asilomar}
S.~Ramprashad, G.~Caire, and H.~Papadopoulos, ``Cellular and network {MIMO}
  architectures: {MU-MIMO} spectral efficiency and costs of channel state
  information,'' in \emph{Proc. IEEE Asilomar Conference on Signals, Systems,
  and Computers.}, November 2009.

\bibitem{ShaZai01}
S.~Shamai and B.~M. Zaidel, ``Enhancing the cellular downlink capacity via
  co-processing at the transmitting end,'' in \emph{Proc. IEEE Veh. Technol.
  Conf.}, vol.~3, 2001, pp. 1745 -- 1749.

\bibitem{SomZaiSha07}
O.~Somekh, B.~M. Zaidel, and S.~Shamai, ``Sum rate characterization of joint
  multiple cell-site processing,'' \emph{IEEE Trans. Inform. Theory}, vol.~53,
  no.~12, pp. 4473 -- 4497, Decemeber 2007.

\bibitem{Qcomm10CTW}
S.~Annapureddy, A.~Barbieri, S.~Geirhofer, S.~Mallik, and A.~Gorokhov,
  ``Coordinated joint transmission in {WWAN},'' in \emph{{IEEE Communication
  Theory Workshop}}, May 2010.

\bibitem{Irmeretal11CommMag}
R.~Irmer, H.~Droste, P.~Marsch, M.~Grieger, G.~Fettweis, S.~Brueck, H.-P.
  Mayer, L.~Thiele, and V.~Jungnickel, ``Coordinated multipoint: Concepts,
  performance, and field trial results,'' \emph{IEEE Commun. Mag.}, vol.~49,
  pp. 102 -- 111, Feburary 2011.

\bibitem{Loveetal08}
D.~J. Love, R.~W.~H. Jr, V.~K.~N. Lau, D.~Gesbert, B.~D. Rao, and M.~Andrews,
  ``An overview of limited feedback in wireless communication systems,''
  \emph{IEEE J. Select. Areas Commun.}, vol.~26, no.~8, pp. 1341--1365, October
  2008.

\bibitem{Somekhetal08}
O.~Somekh, O.~Simeone, A.~Sanderovich, B.~M. Zaidel, and S.~Shamai, ``On the
  impact of limited-capacity backhaul and inter-users links in cooperative
  multicell networks,'' in \emph{42nd Annual Conference on Information Sciences
  and Systems}, 2008.

\bibitem{SanSomPooSha09}
A.~Sanderovich, O.~Somekh, H.~V. Poor, and S.~Shami, ``Uplink macro diversity
  of limited backhaul cellular network,'' \emph{IEEE Trans. Inform. Theory},
  vol.~55, no.~8, pp. 3457 -- 3478, Augest 2009.

\bibitem{FemtoBackhaul}
``3{GPP} {TS} 25.467 v9.3.0: Utran architecture for 3{G} {Home NodeB (HNB)}
  (release 9),'' 3{GPP}, June 2010.

\bibitem{ZhaKouAndHea11Multimode}
J.~Zhang, M.~Kountouris, J.~G. Andrews, and R.~W. {Heath Jr}, ``Multi-mode
  transmission for the mimo broadcast channel with imperfect channel state
  information,'' \emph{IEEE Transactions on Communications}, vol.~59, no.~3,
  pp. 803--814, March 2011.

\bibitem{FosKarVal06}
G.~Foschini, K.~Karakayali, and R.~A. Valenzuela, ``Coordinating multiple
  antenna cellular networks to achieve enormous spectral efficiency,''
  \emph{IEE Proc. Commun.}, vol. 152, no.~4, pp. 548-- 555, August 2006.

\bibitem{HuaWei09}
D.~Wei, ``Leading edge--{LTE} requirements for bearer networks,'' \emph{Huawei
  Communicate}, pp. 49 --51, June 2009.

\bibitem{FleStoSim97}
P.~J. Fleming, A.~L. Stolyar, and B.~Simon, ``Closed-form expressions for
  other-cell interference in cellular {CDMA},'' Technical Report, University of
  Colorado at Denver, December 1997.

\bibitem{AndBacGan10}
J.~G. Andrews, F.~Baccelli, and R.~K. Ganti, ``A tractable approach to coverage
  and rate in cellular networks,'' \emph{Submitted to IEEE Transactions on
  Communications}, September 2010, [Available Online]:
  http://arxiv.org/abs/1009.0516.

\bibitem{DhiGanBacAnd11}
H.~S. Dhillon, R.~K. Ganti, F.~Baccelli, and J.~G. Andrews, ``Modeling and
  analysis of {K}-tier downlink heterogeneous cellular networks,''
  \emph{submitted to IEEE Journal on Sel. Areas in Comm.}, March 2011,
  [Available Online]: http://arxiv.org/abs/1103.2177.

\bibitem{JoSanXiaAnd11}
H.-S. Jo, Y.~J. Sang, P.~Xia, and J.~G. Andrews, ``Outage probability for
  heterogeneous cellular networks with biased cell association,''
  \emph{submitted to IEEE Global Telecommunications Conference}, 2011.

\bibitem{AliTse10}
M.~A. Maddah-Ali and D.~Tse, ``Completely stale transmitter channel state
  information is still very useful,'' in \emph{Allerton Conference on Commun.,
  Control, and Computing.}, September 2010, pp. 1188 -- 1195.

\bibitem{ChaAnd09UL}
V.~Chandrasekhar and J.~G. Andrews, ``Uplink capacity and interference
  avoidance for two-tier femtocell networks,'' \emph{IEEE Transactions on
  Wireless Communications}, vol.~8, no.~7, pp. 3498--3509, July 2009.

\bibitem{GhoZhaAndMuh10LTE}
A.~Ghosh, J.~Zhang, J.~G. Andrews, and R.~Muhamed, \emph{Fundamentals of LTE},
  Englewood Cliffs, New Jersey, 2010.

\bibitem{LTE9}
``3{GPP} {TR} 36.814 v9.0.0: Further advancements for {E-UTRA} physical layer
  aspects (release 9),'' 3{GPP}, March 2010.

\bibitem{VenKul08WirelessBackhaul}
G.~K. Venkatesan and K.~Kulkarni, ``Wireless backhaul for {LTE} - requirements,
  challenges and options,'' in \emph{IEEE International Symposium on Advanced
  Networks and Telecommunication Systems}, Decemeber 2008.

\bibitem{WerWanSyn07}
M.~Wernersson, S.~W$\ddot{a}$nstedt, and P.~Synnergren, ``Effects of {QoS}
  scheduling strategies on performance of mixed services over {LTE},'' in
  \emph{Proc. of the IEEE Int. Symp. on Personal Indoor and Mobile Radio
  Comm.}, 2007.

\bibitem{SadmadSam09}
B.~Sadiq, R.~Madan, and A.~Sampath, ``Downlink scheduling for multiclass
  traffic in {LTE},'' \emph{EURASIP Journal on Wireless Communications and
  Networking}, July 2009.

\bibitem{SA97}
S.~V. Amari and R.~B. Misra, ``Closed-form expression for distribution of the
  sum of independent exponential random variables,'' \emph{IEEE Trans.
  Reliability}, vol.~46, no.~4, pp. 519 -- 522, December 1997.

\bibitem{SF10}
S.~Favaro and S.~G. Walker, ``On the distribution of sums of independent
  exponential random variables via {W}ilks' integral representation,''
  \emph{Acta Applicandae Mathematicae}, vol. 109, no.~3, pp. 1035--1042, March
  2010.

\end{thebibliography}
\newpage
\begin{table}[ht]
\caption{notation $\&$ Simulation Summary} \centering
\begin{tabular}{c|c|c}
\hline \hline Symbol & Description & Simulation Value\\
\hline
\hline $\lambda_1$ & Macro BS density & $5\times 10^{-7}$/$m^2$ (average cell radius of $1$ Km)\\
\hline $\lambda_2$ & Pico BS density & $5 \times 10^{-6}$/$m^2$ (average of $10$ picos/macrocell)\\
\hline $\lambda_3$ & Femto BS density & $5 \times 10^{-5}$/$m^2$ (average of $100$ femtos/macrocell)\\
\hline $P_1$ & Macro BS transmitting power & $40$ W\\
\hline $P_2$ & Pico BS transmitting power & $1$ W\\
\hline $P_3$ & Femto BS transmitting power & $200$ mW\\
\hline $\alpha_1$ & Path loss exponent of Macro BSs &$3.0$\\
\hline $\alpha_2$ & Path loss exponent of Pico BSs& $3.5$\\
\hline $\alpha_3$ & Path loss exponent of Femto BSs & $4.0$\\
\hline $L_w$ & Wall penetration loss (femto BSs are indoor) & $5$ dB \\
\hline $k$ & The tier index of $\mathrm{BS}_n$ & k=1, 2 or 3\\
\hline $q_k\{\cdot\}$ & SIR CDF of wireless overhead channel & N/A\\
\hline $W$ & Wireless channel bandwidth & N/A \\
\hline $N_{IP}$ & Number of servers in IP access network (for femtocells) & 10 \\
\hline $N_{CN}$ & Number of servers in core network & 10 \\
\hline $N$ & Total number of servers in backhaul path &N/A\\
\hline $\bar{\mu}$ & Backhaul servers' average rate (bps) for overhead packets & N/A \\
\hline $B$ & Overhead packet size & $30$ bits \\
\hline $\mathcal{T}$ & Overhead packet interarrival time & N/A \\
\hline $\eta$ & Average overhead packet rate, i.e. $\eta = 1/\mathbb{E} (\mathcal{T})$ & N/A\\
\hline $M$ & Parameter in the distribution of $\mathcal{T}$ &$\mathcal{T} \sim$ Gamma $\left(M, \frac{1}{M\eta}\right)$\\
\hline $d$ & Overhead delay requirement & N/A\\
\hline $\beta(x)$ &SIR target for a given overhead delay requirement $x$& $\frac{B}{W\log (1+\beta(x))}=x$\\
\hline\hline
\end{tabular} \label{table1}
\end{table}


\end{document}